\documentclass[nofootinbib,superscriptaddress,twocolumn,showpacs,preprintnumbers,amsmath,amssymb]{revtex4-1}

\usepackage{graphicx}
\usepackage{grffile}
\usepackage{slashed}
\usepackage{bbm}
\usepackage{footmisc}
\usepackage{multirow}
\usepackage{hyperref}
\usepackage{amssymb}
\usepackage{amsmath}
\usepackage{color}
\usepackage{xspace}
\usepackage{caption}
\usepackage{subcaption}

\usepackage{datetime}
\usepackage{color,fancybox}

\newcommand{\tr}{\mathrm{tr}}

   \def\beq{\begin{equation}} \def\eeq{\end{equation}}
\def\bea{\begin{eqnarray}} \def\eea{\end{eqnarray}}

\begin{document}

\title{Higgs boson production in association with
  three jets via gluon fusion at the LHC: Gluonic contributions}
\preprint{FTUV-13-0606\;\;IFIC-13-27\;\;KA-TP-13-2013\;\; LPN13-034\;\; SFB/CPP-13-38 \;\;TTK-13-16}

\author{Francisco Campanario} 
\email{francisco.campanario@ific.uv.es} \affiliation{Institute for Theoretical
  Physics, KIT, 76128 Karlsruhe, Germany.}
\affiliation{Theory Division,
  IFIC, University of Valencia-CSIC, E-46100 Paterna, Valencia,
  Spain}

\author{Michael Kubocz}
\email{kubocz@physik.rwth-aachen.de} \affiliation{Institute for Theoretical Physics, KIT, 76128 Karlsruhe, Germany.}\affiliation{Institut f\"ur
  Theoretische Teilchenphysik und Kosmologie,\\ RWTH Aachen
  University, D52056 Aachen, Germany}

\begin{abstract} 
  \noindent Higgs production in association with three jets via gluon
  fusion~(GF) is an important channel for the measurement of the
  $\mathcal{CP}$-properties of the Higgs particle at the LHC. In this
  letter, we go beyond the heavy top effective theory approximation
  and include at LO the full mass dependence of the top- and
  bottom-quark contributions. We consider the dominant sub-channel $gg
  \to H ggg$ which involves the manipulation of massive rank-5 hexagon
  integrals. Furthermore, we present results for several differential
  distributions and show deviations from the effective theory as large
  as 100$\%$ at high $p_T$ for light Higgs masses.
\end{abstract}

\pacs{12.38.Bx, 13.85.-t, 14.65.Fy, 14.65.Ha, 14.70.Dj, 14.80.Bn} 
\keywords{Higgs boson, Standard Model, Hadronic Colliders}

\maketitle
\section{Introduction}
\label{intr}
\noindent Higgs production in association with two jets via gluon
fusion is known to be an important channel at the LHC in order to
measure the $\mathcal{CP}$-properties of the new found scalar
particle. Indeed, the differential distribution of the azimuthal angle
between the more forward and the more backward of the two tagging
jets, $\phi_{jj} = \phi_{jF}-\phi_{jB}$, provides a sensitive probe
for the $\mathcal{CP}$-character of the Higgs couplings to
quarks~\cite{Plehn:2001nj, Hankele:2006ma, Klamke:2007cu,
  Hagiwara:2009wt, Campanario:2010mi}. A further aspect of interest is
the modification of the azimuthal angle correlation by emission of
additional jets, that is, at least by a third jet. Former
investigations with showering and hadronization provided a strong
de-correlation between the tagging jets in Higgs plus two jet
production ~\cite{Odagiri:2002nd}. The de-correlation effects,
however, were disproportionately illustrated due to approximations in
the parton-shower. Further
analyses~\cite{DelDuca:2006hk,DelDuca:2008zz,Andersen:2010zx} also
showed, that after the separation of the hard radiation from the showering
effects with subsequent hadronization, the $\phi_{jj}$-correlation survives
with minimal modifications. Similar results were obtained by a parton
level calculation with NLO corrections~\cite{Campbell:2006xx} to the
Higgs plus two jets process in the framework of an effective
Lagrangian.
\noindent In this letter, we provide results for the sub-process $gg
\to H ggg$ going beyond the heavy top approximation, including the
full mass dependence of the top- and bottom-quark contributions at
LO. This sub-process involves the manipulation of massive rank-5
hexagon Feynman diagrams, which are the most complicated topologies
appearing in Higgs production in association with three jets via GF,
and thus it provides a testing ground to check the numerical stability
of the full process.
This is particularly important for the numerically challenging
bottom-loop corrections, which are small within the SM, however, once
a $\mathcal{CP}$-odd Higgs is considered, large corrections can arise
for large values of $\tan \beta$, which can be used to discriminate
the $\mathcal{CP}$-properties of the new found scalar particle-- a
pure $\mathcal{CP}$-odd scalar Higgs has been already discarded with
more than three standard deviations~\cite{Freitas:2012kw}, however, a
$\mathcal{CP}$-violating Higgs boson consisting of a mixture of
$\mathcal{CP}$-odd and $\mathcal{CP}$-even couplings to fermions, is
still not.
Within the SM, $gg \to H ggg$ is the dominant channel, hence, an
essential piece to compute the real emission contributions for Higgs
plus two jets production at NLO via GF. Results for the full process
and a detailed description of de-correlation effects will be given in
a forthcoming publication.
\noindent Furthermore, the presented results are important to study
the validity of the effective Lagrangian approach in Higgs production
plus two and three jets. They are also relevant for heavy-Higgs
searches beyond the Standard Model, since large deviations with
respect to the effective theory are expected for a Higgs mass,
$m_{H}$, bigger than twice the top-quark mass, $m_t$.

\noindent This letter is organized as follows: in
Section~\ref{calculations}, the details of the calculations are
given. Numerical results are presented in
Section~\ref{results}. Finally, we summarize in Section~\ref{sec:end}.


\label{calculations}
\begin{figure}[!ht]
\begin{center}
  \includegraphics[width=1\columnwidth]{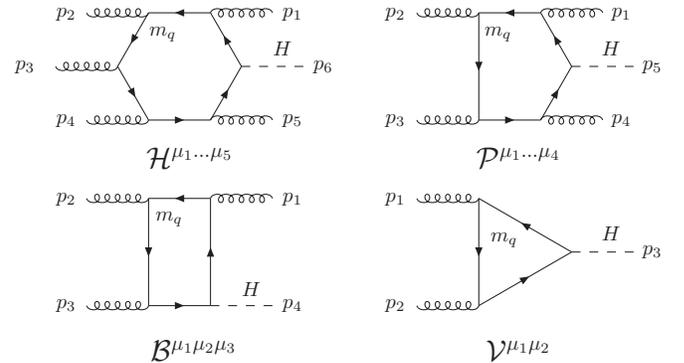}\quad
\vspace*{-1.5em}
\end{center}
\caption{Master Feynman diagrams}
\label{fig:diag}
\end{figure}
\section{Calculation}
\label{calculations}
\noindent The production of the $\mathcal{CP}$-even Higgs boson $H$ at
order $\alpha_s^5$ can be carried out via the following sub-processes
(+ crossing related)~\cite{Kubocz:2009PhD}:
\begin{align}  \label{3jsubproc}
q\, q \rightarrow q\, q\,g\, H , \quad q\, Q \rightarrow q\,
Q\,g\, H , \nonumber \\ 
q\, g \rightarrow q\, g\,g\, H , \quad
g\ g \rightarrow g\, g\, g\,H \, .
\end{align}
In this letter, we restrict our analysis to the last sub-channel containing only
gluons in the initial and final state. We use the effective current
approach~\cite{Hagiwara:1985yu, Hagiwara:1988pp}, which allows us to
compute only four master Feynman diagrams, depicted in
Fig.~\ref{fig:diag}. Note that the attached gluons 
are treated generally as off-shell vector currents. The calculation of the master integrals is
performed with the in-house framework described in
Ref.~\cite{Campanario:2011cs}. 
Additionally, we reduce the number of diagrams to be computed 
applying the Fury's theorem to all contributing topologies.
Based on the gluon fusion part {\em GGFLO} of the program {\em
  VBFNLO}~\cite{Arnold:2008rz}, we devote special care to the
development of a fully-flexible, numerically stable parton-level Monte
Carlo program.
The control of numerical instabilities, appearing in each of the
master diagrams due to vanishing Gram determinants, is done 
evaluating Ward identities, which replace the polarization vectors of attached
gluons by their corresponding momenta. This allows us to relate and
additionally check the master Feynman diagrams depicted in
Fig.~\ref{fig:diag}. For example, a hexagon topology of rank five is
written as a difference of two pentagons topologies of rank four
\begin{equation}
{\cal H}^{\mu_1 \ldots \mu_5} p_{i,\mu_i} =
{\cal P}_1^{\mu_1 \ldots \hat{\mu}_i \ldots \mu_5}
- {\cal P}_2^{\mu_1 \ldots \hat{\mu}_i \ldots \mu_5},
\hspace{0.05cm} i  =1\ldots 5 \,,
\label{ward}
\end{equation}
where $\hat{\mu}_i$ denotes the corresponding vertex replaced by its
momentum $p_i$.
We construct all possible Ward identities for each physical permutation
and diagram, e.g.\ all five different ones for the hexagon
${\cal H}^{\mu_1\ldots \mu_5}$. 
These Ward identities are evaluated for each phase space point and
diagram with a small additional computing cost using a cache
system. If the identities are not satisfied better than five per ten
thousand level for a given diagram, this one is reevaluated by
computing the scalar integrals and tensor reduction routines in
quadruple precision. The complete phase-space point is rejected and
the amplitude set to zero if the Ward Identities are not satisfied
after this step. With this system, we find that the amount of
phase-space points, which does not pass the Ward identities for a
requested accuracy of $\epsilon=5 \times 10^{-4}$, is statistically
negligible and well below the per mille level, taking into account
also the numerically challenging bottom-loop contributions (5$\%$ of the phase
space points are rejected when using only double precision). This
method was also applied successfully in ZZ+jet production via GF in
Ref.~\cite{Campanario:2012bh}. Final results are given demanding a
global accuracy of the Ward identities of $\epsilon=5 \times 10^{-4}$.
For the numerical evaluation of tensor integrals, we apply the
Passarino-Veltman approach of Ref.~\cite{Passarino:1978jh} up to
boxes, and for a numerically stable implementation of
five-point-coefficients, we use the  scheme laid out in 
Ref.~\cite{Campanario:2011cs}. Corresponding color factors were
computed by hand and cross-checked with the program
\textsc{Madgraph}~\cite{Alwall:2007st,Alwall:2011uj}.
To define a color basis, it is strategically favorable to start with
hexagons and investigate their color structure. The five external
gluons give rise to $5!=120$ hexagons (60 after applying Fury's
theorem) proportional to color traces of the form
\begin{align}
  \tr \big[ t^{a_i} t^{a_j} t^{a_k} t^{a_l} t^{a_m} \big] \quad
  &\text{with} \quad i,j,k,l,m=1,\ldots,5 \nonumber \\
  &\text{and} \quad i \neq j \neq k \neq l \neq m~,
\end{align}
in which $(5-1)!=24$ are independent of each other. Thus, they can be
used to form a color basis for all remaining amplitudes with
triangle-, box- and pentagon-like topologies.

\noindent To cross check our results, we have compared the top-loop
contribution with the heavy top-mass approximation, which is also a
part of the GF-implementation within the VBFNLO framework.
The agreement at the integrated cross section level for
$m_t=5\cdot 10^4$ GeV is better than one per ten
thousand. We have also performed a comparison with Madgraph
and got agreement at the per mille level. Additionally, gauge invariance was
checked
 at the amplitude level with expected cancellations of the order of the machine precision.
\begin{figure*}[!ht]
\includegraphics[width=1.0\columnwidth]{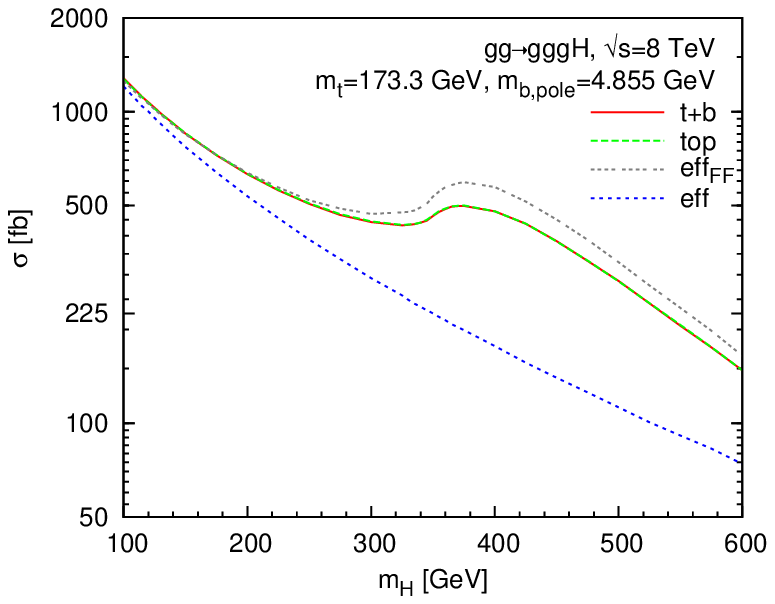}
\includegraphics[width=1.0\columnwidth]{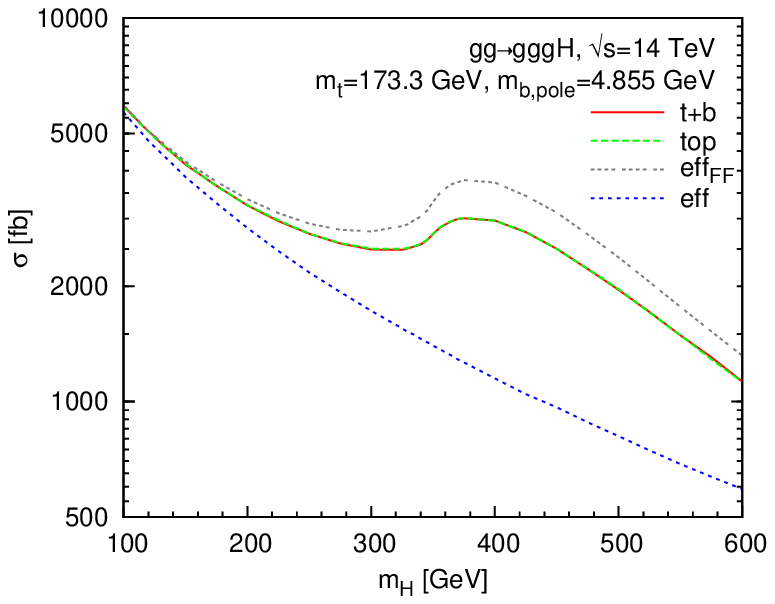}
\caption{Cross section of the $gg \rightarrow gggH$ scattering
  sub-process as a function of Higgs boson mass for c.o.m. energies of
  8 TeV and 14 TeV. Both panels show the effective theory (with and
  without FF) and full mass contributions. The inclusive cuts (IC) of
  Eq.~\eqref{ICuts} have been applied.}
\label{Hmass-scan}
\end{figure*}
\section{Numerical Results}
\label{results}
\noindent In this section, we present integrated cross sections and
differential distributions for the sub-process $gg \to ggg H$ at the
LHC for various center of mass (c.o.m.) energies.
The Higgs boson is produced on shell and without finite width effects.  We set
the top-quark mass to $m_t=173.3$\,GeV, the $\overline{\text{MS}}$
bottom-quark mass to $\overline{m}_b(m_b)=4.2$\,GeV, and the other light quark
masses to zero. Within the Higgs-mass range of 100-600 GeV, the bottom-quark
mass is 33-42$\%$ smaller than the pole mass of 4.855 GeV used in the loop
propagators. Thus, we take into account the evolution of $m_b$ up to a
reference scale, here $m_H$, and the relation between the pole mass and the
$\overline{\text{MS}}$ mass. Additionally, we choose $ M_Z = 91.188
\text{GeV}$, $M_W=80.386 \text{GeV}$ and $G_F=1.16637\times 10^{-5}
\text{GeV}^{-2}$ as electroweak input parameters and derive further necessary
parameters from Standard Model tree level relations.

\noindent Cross section predictions are obtained using the CTEQ6L1
parton distribution functions (PDFs) \cite{Pumplin:2002vw} with the
default strong coupling value $\alpha_s(M_Z)=0.130$.
The factorization scale is set to
$\mu_F=(p_{T}^{j1} p_{T}^{j2} p_{T}^{j3})^{1/3}$ and the
renormalization to
\begin{equation}
  \alpha_s^{5}(\mu_R) = \alpha_s(p_T^{j_1}) \alpha_s(p_T^{j_2})
  \alpha_s(p_T^{j_3}) \alpha_s(p_H)^2
  \, .
\end{equation}
Here, $p_T^{j_i}$ with $i=1,2,3$ denotes jets with decreasing
transverse momenta. We use the $k_T$ jet algorithm and impose
\begin{align} \label{ICuts}
p_{T}^{j_i} > 20 \ \text{GeV}\;, \qquad |\eta_j| < 4.5\;, \qquad R_{jj} > 0.6 \; ,
\end{align}
where $R_{jj}$ describes the separation of the two partons in the
pseudo-rapidity versus azimuthal-angle plane,
\begin{align}
R_{jj} = \sqrt{\Delta \eta_{jj}^2 + \phi_{jj}^2} \;,
\end{align}
with $\Delta \eta_{jj} = |\eta_{j1}-\eta_{j2}|$ and $\phi_{jj} =
\phi_{j1}-\phi_{j2}$. These cuts anticipate LHC detector capabilities
and jet finding algorithms and will be called ``inclusive cuts'' (IC).
\begin{table}
\begin{tabular}{|c||c|c|c|c|}
\hline
$m_H$ &$\sigma$[fb] / $\sqrt{s}$ &7 TeV &8 TeV & 14 TeV \\\hline
\hline
\multirow{2}{*}{126 GeV} & $\sigma(\text{t})$ & 674.7 $\pm $ 0.3 &
1017 $\pm $ 0.5 &  4846 $\pm $ 1 \\\cline{2-5}
& $\sigma(\text{t+b})$ &   676 $\pm $ 0.4 & 1019 $\pm $ 0.5 & 4864 $\pm $ 1 
\\ \hline
\hline
\multirow{2}{*}{400 GeV} & $\sigma(\text{t})$ & 292.9 $\pm $ 0.2&
480.2 $\pm $ 0.3 &  2965 $\pm $ 0.8  \\\cline{2-5}
& $\sigma(\text{t+b})$& 292.6 $\pm $ 0.2&480.1 $\pm $ 0.3 & 2962 $\pm $ 1
\\ \hline
\end{tabular}
\caption{Cross sections evaluated for different Higgs masses and c.o.m.
  energies applying the inclusive  set of cuts (IC) of
  Eq.~\eqref{ICuts}.}
\label{cstable}
\end{table}
\noindent Values of cross sections for two different Higgs masses
evaluated at different c.o.m. energies are summarized in
table~\ref{cstable}.  A Higgs mass of 400 GeV has been chosen to show maximal
deviations of the effective theory approximation for Higgs
masses larger than $ 2 m_t$ despite of the experimental SM Higgs boson exclusion limits.
As expected, bottom-loop corrections hardly
contribute to the overall cross section, and hence, they can be
neglected within the SM framework. Noticeable is the negative impact
of the interference term between top- and bottom-loop induced
contributions for a Higgs mass of 400 GeV, which decreases the overall
cross section by a small amount.

\noindent The total cross section as a function of the Higgs boson
mass is shown in Fig.~\ref{Hmass-scan} for c.o.m. energies of 8 TeV
(left panel) and 14 TeV (right panel). Amplitudes with a top-loop mediated
contribution give rise to a striking peak due to the threshold
enhancement around $m_H \approx 2m_t$. Here, we removed the singular
behavior at $m_H = 2m_t$ by omitting the corresponding phase-space
point. For the effective theory approximation, we used two approaches:
pure effective Higgs coupling to fermions in the heavy top-quark limit
with and without corrections by an additional form factor
(FF)~\cite{Djouadi:2005gi}. Up to Higgs masses of 150 (175) GeV (within 10
\% deviation with respect to the full theory)
, the effective theory approximation
gives accurate results for a c.o.m. energy of 8 (14) TeV. 
%
The application of the form factor FF extends additionally the
validity range of the effective approximation to higher Higgs masses,
here up to 330 GeV for $\sqrt{s}=8$ TeV (within $10 \%$ deviation with
respect to the full theory), whereas for $\sqrt{s}=14$ TeV the upper
validity bound is fixed at 290 GeV. Furthermore, it gives back the
top-quark mass dependence imitating the threshold enhancement of the
full theory. Beyond the validity bound, the total cross section is
overestimated up to 20~(25)$\%$ for 8~(14) TeV c.o.m. energy at $m_H=400$ GeV,
and converges afterwards slowly to the full theory result for the shown range of Higgs mass.
\begin{center}
\begin{figure*}[!thb]
\includegraphics[width=0.957\columnwidth]{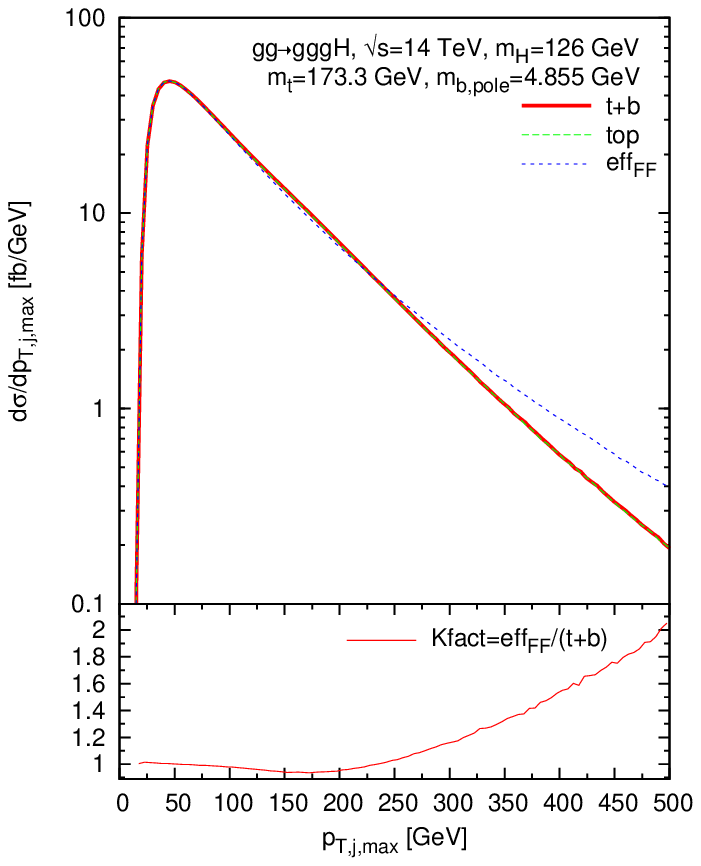}
\hfill
\includegraphics[width=0.957\columnwidth]{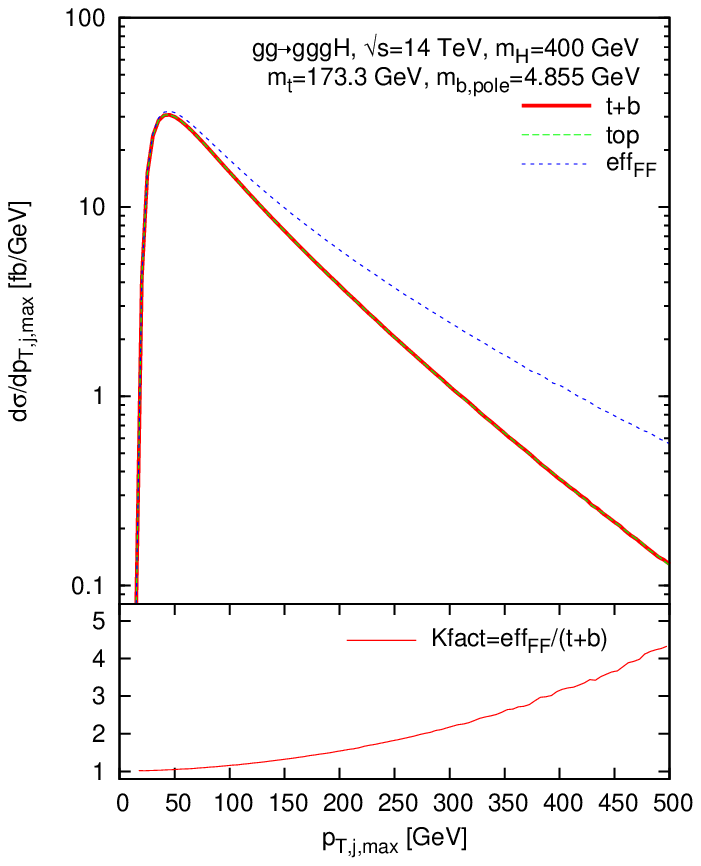}
\caption{Comparison of transverse-momentum distributions of the harder
  jet of the $gg \rightarrow gggH$ scattering sub-process evaluated
  within the effective and loop-induced theory. The inclusive cuts
  (IC) of Eq.~\eqref{ICuts} are applied.}
\label{pthjet:diff}
\end{figure*}
\begin{figure*}[!thb]
\includegraphics[width=0.957\columnwidth]{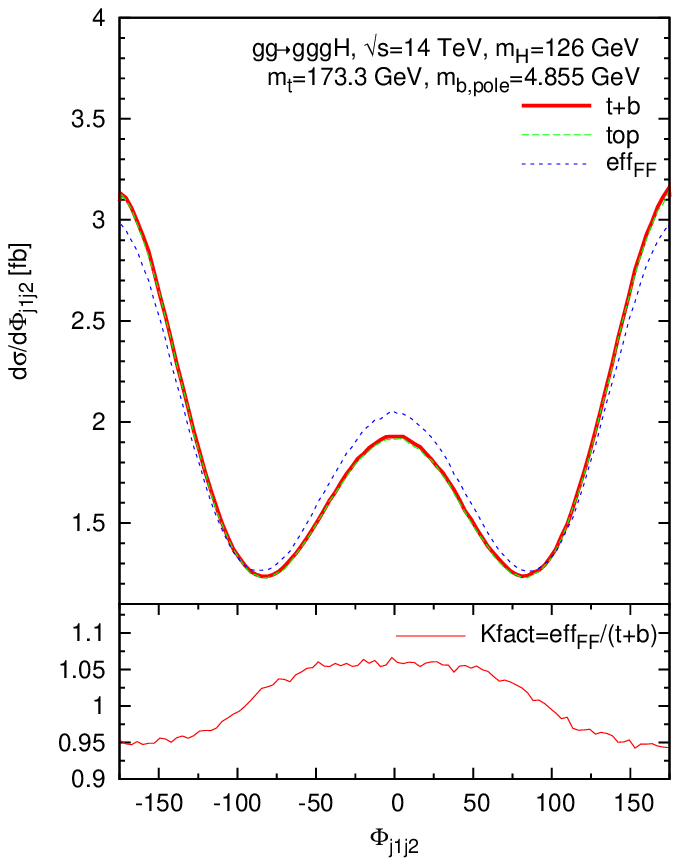}
\hfill
\includegraphics[width=0.957\columnwidth]{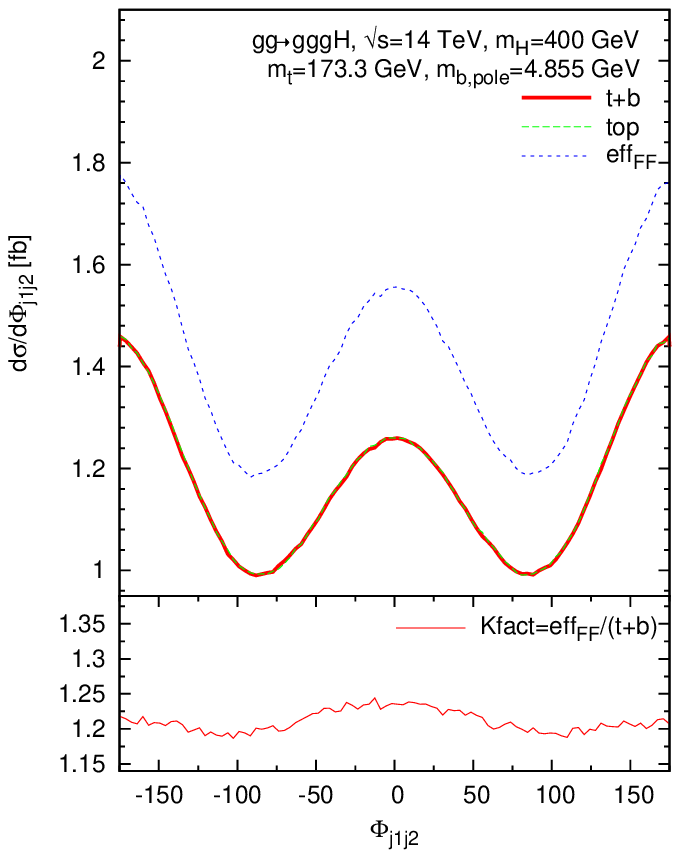}

\caption{Comparison of azimuthal angle distributions of the $gg
  \rightarrow gggH$ scattering sub-process evaluated within the
  effective and loop-induced theory. The ICphi set of acceptance cuts
  of Eq.~\eqref{ICphi} is applied.}
\label{azim:diff}
\end{figure*}
\end{center}
\noindent Next, for a c.o.m. energy of 14 TeV, we present some
differential distributions for Higgs masses of 126 GeV (left panels)
and 400 GeV (right panels).  Presented results were evaluated with
contributions mediated by quark-loops, $(t+b)$, and with the effective
theory framework including form factor corrections,
eff$_{\text{FF}}$. Differences between both approaches are illustrated
with the help of a $K$-factor defined as
$\text{eff}_{\text{FF}}/(t+b)$. The differential distributions for the
hardest jet are shown in Fig.~\ref{pthjet:diff}. In the left panel,
one can see that the effective theory provides, as expected, a very
good approximation of the full theory up to $p_{T}^{j\text{max}} <
200$ GeV. \linebreak[4] \\\\Beyond that regime, differences start to increase and
deviations up to $100\%$ are found. The discrepancy of the
differential distribution illustrated in the right panel are more
evident. At the maximum of the differential distribution, the
deviation of the effective theory is only about 5$\%$. As the $p_T$
increases, the effective theory predicts harder emissions, which are
overestimated up to a factor of 5 and add to the total 25$\%$
discrepancy at the level of the total cross section.

\noindent Following the definition of Ref.~\cite{Hankele:2006ma}, the
azimuthal angle distributions between the more forward and the more
backward of the two tagging jets are depicted in
Fig.~\ref{azim:diff}. The calculation was carried out with a modified
set of cuts, which leads to a better sensitivity to the
$\mathcal{CP}$-structure of the Higgs couplings than the inclusive
cuts. We use
\begin{align} 
\label{ICphi} p_{T}^{j_i} > 30 \ \text{GeV}\;, \hspace{0.1cm} |\eta_j| < 4.5\;,
\hspace{0.1cm} R_{jj} > 0.6\;, \hspace{0.1cm} \Delta \eta_{jj}>3 \;,
\end{align}
and label it as the ICphi set of cuts in the following.

\noindent The effective theory approximation provides a good
des\-crip\-tion in the whole azimuthal angle range with a deviation up
to the 5$\%$ level for a 126 GeV massive Higgs boson. 
For a 400 GeV Higgs boson mass, the shape is well reproduced, but 20 \% off
in the whole spectrum.
%
%
\section{Summary}
\label{sec:end}
\noindent We have presented a short analysis of the gluon fusion
loop-induced sub-process $gg \to ggg H$ at the LHC, which is the
dominant one in Higgs production in association with three jets via
GF. We devoted special attention to the development of a numerical
stable MC program which solves the problem of vanishing Gram
determinants by suitable application of Ward identities and quadruple
precision. We have also included bottom-loop corrections, although
they are negligible in the SM framework, to show additionally the
numerical stability of the contributing integrals for small
loop-masses (they will become relevant in combination with a
$\mathcal{CP}$-odd Higgs boson and large $\tan \beta$ values).  Up to
Higgs masses of 290 GeV for $\sqrt{s}=14$ TeV (330 GeV @ 8 TeV) and
for small transverse momenta $p_{T}^{j\text{max}} \lesssim 290$ GeV,
the effective Lagrangian approximation with the form factor correction
gives accurate results and can be used as a numerically fast
alternative for phenomenological studies. No restriction was found in
the validity of the invariant mass of the dijet system of the leading
jets (not shown). For a 400 GeV Higgs mass, large deviation can be
found for small transverse momenta $p_{T}^{j\text{max}} \lesssim
m_t$. Furthermore, the azimuthal angle distribution, sensitive to the
$\mathcal{CP}$-property measurements, is (relatively) well described
by the effective theory both in shape and normalization for (heavy)
light Higgs masses. A detailed description of the full process and
de-correlation effects will be given in a forthcoming publication. 
This process will be made publicly available as
part of the VBFNLO program.
%
\section*{Acknowledgments} 
\noindent It is a pleasure to thank Dieter Zeppenfeld for fruitful
discussions during the development of this project.
We acknowledge the support from the Deutsche Forschungsgemeinschaft
via the Sonderforschungsbereich/Transregio SFB/TR-9 Computational
Particle Physics. FC is funded by a Marie Curie fellowship
(PIEF-GA-2011-298960) and partially by MINECO (FPA2011-23596) and by
LHCPhenonet (PITN-GA-2010-264564) and acknowledges the Institute for
Theoretical Physics, at KIT for the use of the computer
Grid Cluster.
MK acknowledge support by the Grid Cluster of the RWTH-Aachen.
\appendix
\bibliographystyle{h-physrev}
\bibliography{ggh}

\end{document}